\documentclass[aps,prl,notitlepage,10pt,twocolumn,longbibliography,superscriptaddress]{revtex4-1}
\usepackage{bm}
\usepackage{amsmath}
\usepackage{graphicx}
\usepackage{xcolor}
\usepackage{isotope}
\usepackage{hyperref}
\newcommand{\affA}{Physikalisches Institut, Universit\"at Heidelberg, Im Neuenheimer Feld 226, 69120 Heidelberg, Germany.}
\newcommand{\affB}{IPCMS (UMR 7504) and ISIS (UMR 7006), Universit\'e de Strasbourg and CNRS, 67000 Strasbourg, France.}

\begin{document}

\title{Diffusive to non-ergodic dipolar transport in a dissipative atomic medium}

\author{S. Whitlock}\email{whitlock@ipcms.unistra.fr}\affiliation{\affA}\affiliation{\affB}
\author{H. Wildhagen} \affiliation{\affA}
\author{H. Weimer}\affiliation{Institut f\"ur Theoretische Physik, Leibniz Universit\"at Hannover, Appelstra\ss e 2, 30167 Hannover, Germany}
\author{M. Weidem\"uller}\affiliation{\affA}\affiliation{Hefei National Laboratory for Physical Sciences at the Microscale and Department of Modern Physics, and CAS Center for Excellence and Synergetic Innovation Center in Quantum Information and Quantum Physics, University of Science and Technology of China, Hefei, Anhui 230026, China.}

\date{\today}

\begin{abstract}
We investigate the dipole mediated transport of Rydberg impurities through an ultracold gas of atoms excited to an auxiliary Rydberg state. In one experiment we continuously probe the system by coupling the auxiliary Rydberg state to a rapidly decaying state which realizes a dissipative medium. In-situ imaging of the impurities reveals diffusive spreading controlled by the intensity of the probe laser. By preparing the same density of hopping partners but then switching off the dressing fields the spreading is effectively frozen. This is consistent with numerical simulations which indicate the coherently evolving system enters a non-ergodic extended phase due to disorder. This opens the way to study transport and localization phenomena in systems with long-range hopping and controllable dissipation.
\end{abstract}

\maketitle
The transport of charge, energy or information plays a fundamental role in science and technology, for example, determining the function of nanoelectronic devices~\cite{datta2017}, photochemical and biophysical processes~\cite{Scholes2011}, and even the dynamics of complex networks~\cite{mulken2011}. Usually however the relevant transport mechanisms depend very strongly on the underlying structure of each system and its coupling to the environment. For example, in disordered systems governed by short-range hopping, transport can be exponentially suppressed due to Anderson localization~\cite{anderson1958,Lee1985} or many-body localization~\cite{Basko2006,Gornyi2005}. In contrast, long-range hopping or decoherence introduced by coupling to a dissipative environment tends to destroy localization effects. Comparatively little is known about the interface between short-range and long-range hopping (e.g. dipolar $1/r^3$ hopping in three-dimensions)~\cite{Eisert2013,Yao2014,Deng2016,Eldredge2017,Deng2018,alvarez2015} or systems situated at the quantum-classical crossover~\cite{plenio2008,rebentrost2009,chin2010,Bettles2017} which are of special interest for discovering and understanding new transport mechanisms.
 
\begin{figure}[t!]
\includegraphics[width=\columnwidth]{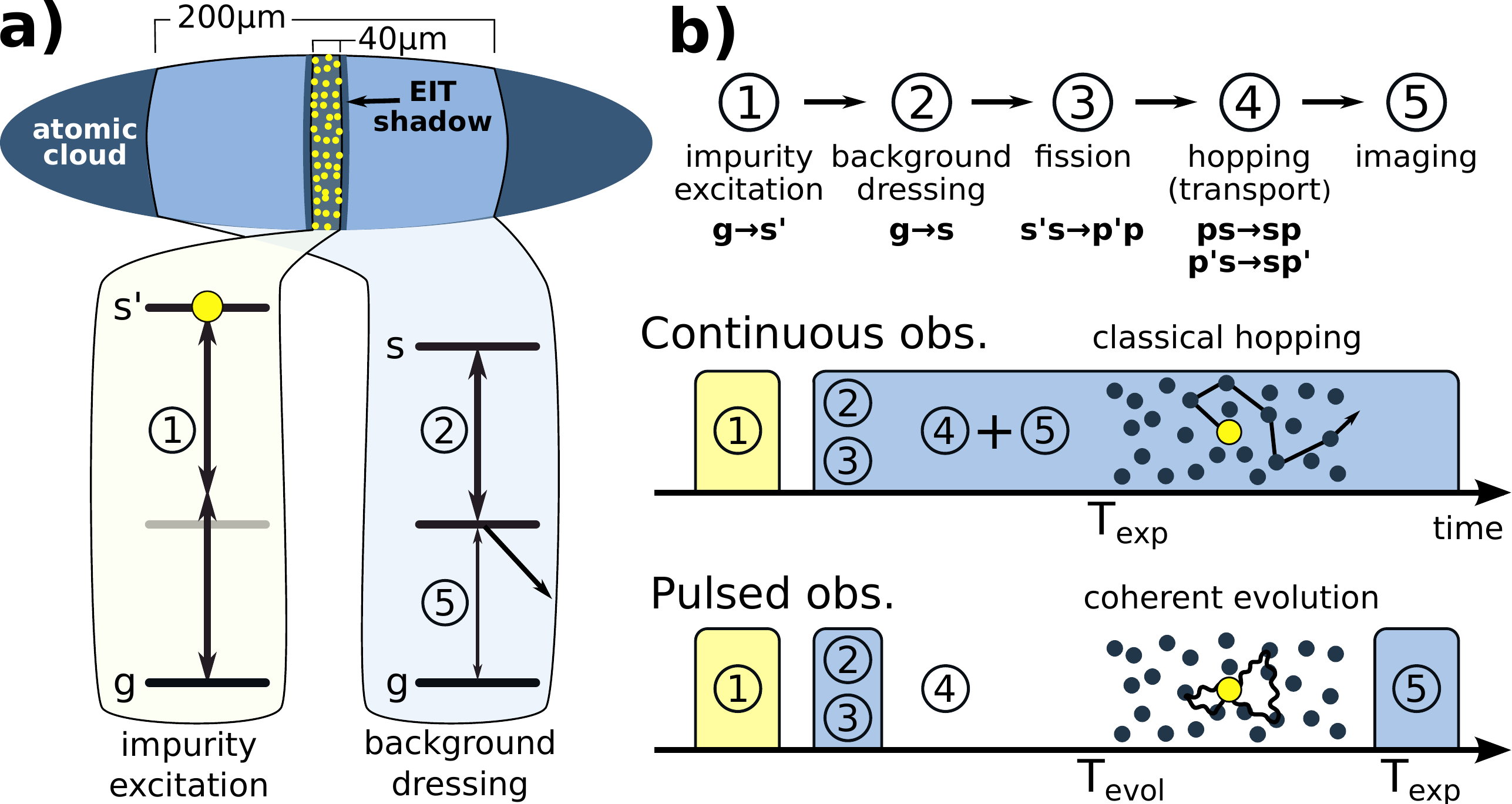}
 \caption{(Color online) Setup for studying dipolar mediated excitation transport in a controllable dissipative medium. (a) Sketch of the experimental geometry and level scheme used to prepare the Rydberg impurities and background gas and to observe transport. (b) The experimental sequence starts exciting a small number of impurities at the center of an ultracold atomic gas. We then couple a larger volume of atoms to a second Rydberg state via an EIT resonance. The presence of impurities is signaled by increased absorption on the probe laser. Different transport behavior is observed depending on whether the probe laser is left on continuously or pulsed.
}
 \label{fig1}
\end{figure}
In this letter we report an experimental and theoretical study of the transport of atomic excitations in a three-dimensional ultracold atomic gas governed by dipolar hopping interactions and subject to a controllable dissipative environment. We start with a small number of Rydberg excitations in a Rydberg $np$ state (``impurities'') embedded in a gas of atoms prepared in an auxiliary Rydberg $ns$ state (``background''). The huge dipole moments of highly-excited Rydberg states leads to coherent energy exchange between $np$ and $ns$ states allowing the impurities to migrate through the gas of $ns$ atoms that acts as a medium~\cite{anderson1998,mourachko1998,muelken2007,scholak2011b,wuester2011,robicheaux2014,schoenleber2015,schempp2015,anderson2002,westermann2006,vanditzhuijzen2008,guenter2013,ravets2014,barredo2014}.  By additionally coupling the $ns$ states to short-lived spontaneously decaying states via laser fields we can spatially resolve the impurity atoms~\cite{guenter2012,guenter2013}. As we show, fundamentally different transport behavior is observed when these lasers are on continuously (thereby continuously watching the dynamics) or when the system is allowed to evolve in the absence of laser fields. This realizes a platform for studying and controlling transport and localization phenomena in classical and quantum regimes, complementary to other experimental platforms involving ground-state atoms~\cite{akkermans2008,roati2008,billy2008,kondov2011,krinner2015,semeghini2015,schreiber2015,choi2016}, trapped ions~\cite{jurcevic2014,richerme2014,smith2016,Trautmann2018}, photonic networks~\cite{viciani2015,biggerstaff2016,harris2017} and superconducting qubits~\cite{mostame2012,potovcnik2018}, with excellent control concerning disorder, dimensionality, dissipation and long-range interactions.
 
We begin our experiments by preparing $10^4$ \isotope[87]{Rb} atoms in a cigar shaped optical trap with a temperature of $40$\,$\mu$K (Fig.~\ref{fig1}a). The cloud has an elongated Gaussian shape with $e^{-1/2}$ radii of $\{12\,\mathrm{\mu m},12\,\mathrm{\mu m},200\,\mathrm{\mu m}\}$ and a peak density of ground state atoms of $0.1~\mu$m$^{-3}$. On the relevant timescale of our experiments (a few tens of microseconds) the atoms are effectively frozen in space such that dynamics only occurs in the internal (Rydberg)-state degrees of freedom. 
The basic processes involved to study transport in our experiment are illustrated in Fig.~\ref{fig1}b. We first create a small number of impurity Rydberg excitations in the $|n's=50s_{1/2}\rangle$ state by switching off the optical trap and then applying a two-photon laser excitation which illuminates a strip of approximately $40\,\mu$m width at the center of the cloud (Fig.~\ref{fig1}a). Following this we couple a larger region of approximately $200\,\mu$m width with a second set of lasers on an electromagnetically induced transparency (EIT) resonance in ladder configuration $|5s_{1/2}\rangle \rightarrow |5p_{3/2}\rangle \rightarrow |ns=48s_{1/2}\rangle$. This involves a strong coupling laser (upper transition) and a weak probe laser (lower transition) for which the transmitted light is additionally resolved on a CCD camera. 

In the absence of impurities, the EIT coupling renders the cloud partly transparent and establishes a small steady-state fraction of Rydberg excitations in the $|ns\rangle$ state, corresponding to a mean inter-Rydberg distance of $8.5\,\mu$m. 
The specific Rydberg states used were chosen because of the near degeneracy of the $|ns,n's\rangle \leftrightarrow |np=48p_{1/2},n'p=49p_{1/2}\rangle$ pair states. As a result, within approximately 1\,$\mu$s each impurity excitation is converted into two $np,n'p$ excitations which can then migrate through the remaining background gas via resonant exchange processes such as $|ns,np\rangle \leftrightarrow |np,ns\rangle$ and $|ns,np'\rangle \leftrightarrow |np',ns\rangle$. This conversion process is very similar to singlet fission processes commonly encountered in molecular photophysical processes~\cite{Smith2013}. In our experiments the $ns,np$ and $ns,n'p$ interactions are almost the same (interaction coefficient $C_3/2\pi\approx $\,5.2\,GHz$\mu$m$^3$) and as such we do not distinguish between $np$ or $n'p$ impurities. 

\begin{figure}[t!]
\includegraphics[width=0.9\columnwidth]{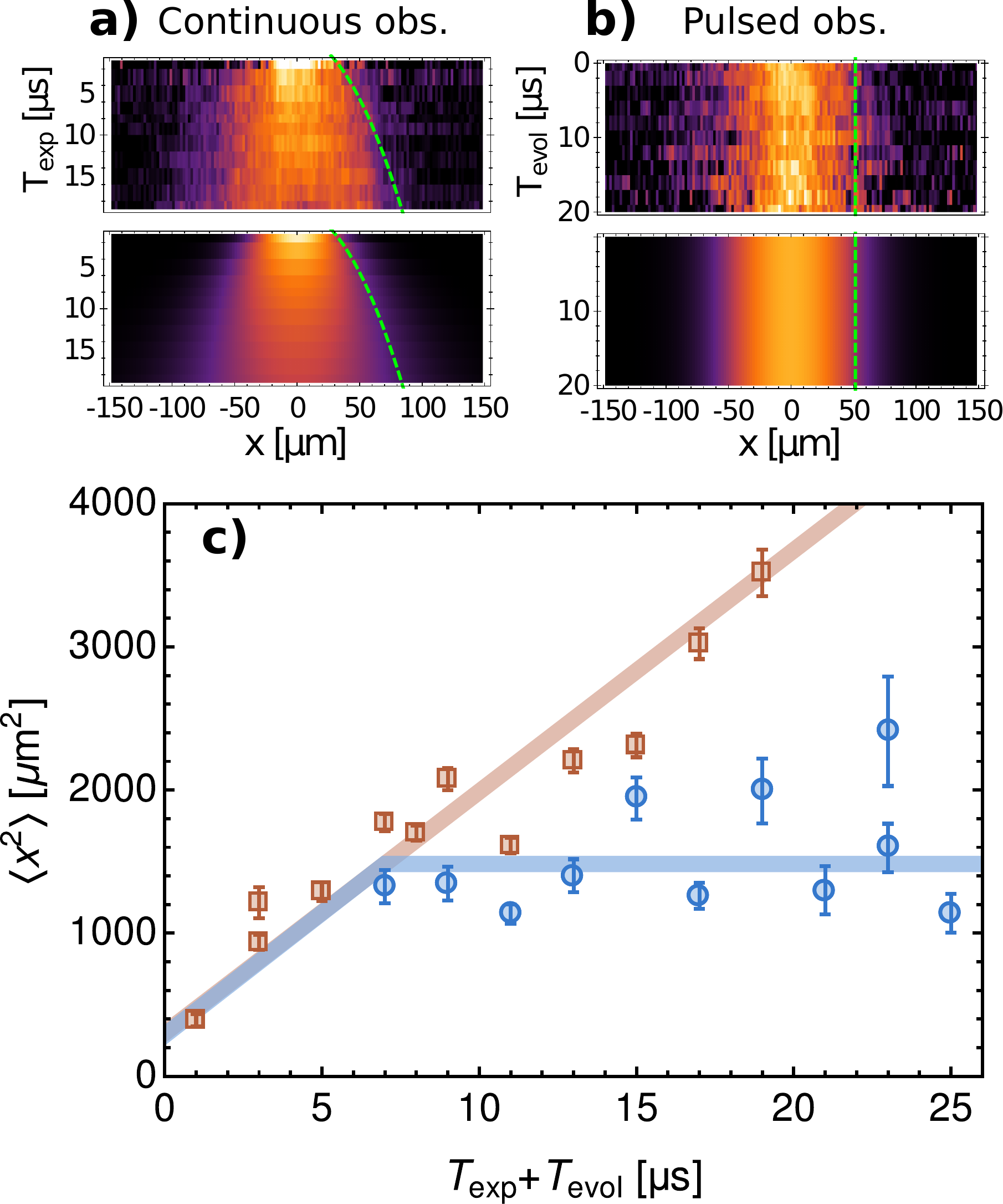}
 \caption{Transport of Rydberg impurities under continuous and pulsed observation. (a) Density plots of the spatial distribution of impurities along the axial $x$ direction as a function of time. The upper panel shows the experimental measurements, the lower panel corresponds to the solution to the classical diffusion law. (b) Corresponding density plots for the case of pulsed observation showing the absence of diffusion. The lower panel is a Gaussian distribution determined by the preparation and probe periods with no dynamics during the evolution period. The dotted green lines in (a) and (b) depict the time-dependent widths of the distribution from the respective time-evolution model. (c) Measured mean square deviation $\langle x^2 \rangle$ as a function of the total time which is the sum of the free evolution time and the exposure time for imaging. The solid lines are fits to the diffusion law showing diffusive (linear) and frozen (constant) dynamical evolution. Each data point is obtained from the average over 100 experimental runs while the error bars represent one standard error of the mean.}
 \label{fig2}
\end{figure}

To monitor the ensuing dynamics we make use of the interaction-enhanced-imaging technique which uses the background atoms as an amplifying medium for detecting impurity atoms~\cite{guenter2012,guenter2013}. In the absence of impurities, the atomic gas is mostly transparent for the probe laser due to EIT. However, in the vicinity of each impurity, the strong dipolar interactions shift the excited state out of resonance thereby breaking the EIT condition for the background gas atoms and locally increasing the absorption. By subtracting images of the probe light taken with and without impurities we recover a two-dimensional image of their spatial distribution. For analysis we integrate this signal over the radial coordinate of the cloud resulting in one-dimensional distributions shown in Fig.~\ref{fig2}a,b. 

To explore both incoherent and coherent transport regimes we perform two different types of experiments. In experiment (1) the probe laser is kept on continuously leading to a continuous spatial measurement of the impurity distribution (which is integrated in time on the CCD camera). In experiment (2) we use a pulsed observation scheme in which we trigger the dynamics with a short probe pulse to prepare a medium with the same density of $|s\rangle$ excitations and then let the system evolve without probe light for a variable time. Afterwards a second probe pulse is applied to spatially resolve the final distribution of impurity atoms.

\noindent\emph{Experiment (1) - Continuous observation}. The continuous spatial projection of the positions of the impurities due to continuous observation leads to transport described by classical diffusion. To demonstrate this, Fig.~\ref{fig2}a shows a density plot of the radially integrated impurity distribution as a function of time. This is compared to a solution to the classical diffusion law $\dot n(x,t)=D\partial^2n(x,t)/\partial x^2$. The experiment clearly shows that the impurities spread through the atomic medium, with the spatial distribution increasing in width by a factor of three within $15\,\mathrm{\mu s}$. The experimentally determined distributions are in excellent agreement with the solution to the diffusion law for an initial Gaussian density distribution with fitted width $\sigma(t=0)=17(1)\,\mathrm{\mu m}$ and diffusion constant $D=84(3)\mu$m$^2$/$\mu$s (also shown in Fig.~\ref{fig2}a), where the uncertainties indicated in parentheses are obtained by bootstrap resampling of the full spatio-temporal dataset for fitting. Taking into account that the signal is averaged over the variable exposure time on the camera, the true diffusion rate is assumed to be a factor of two larger due to temporal averaging~\cite{guenter2013}. The diffusive nature of the transport is further evident in the second central moment $\langle x(t)^2\rangle=\int x^2 n(x,t) dx/\int n(x,t)dx$ of the measured impurity distribution which grows linearly in time (squares in Fig.~\ref{fig2}c).

\begin{figure}[t!]
\includegraphics[width=0.4\textwidth]{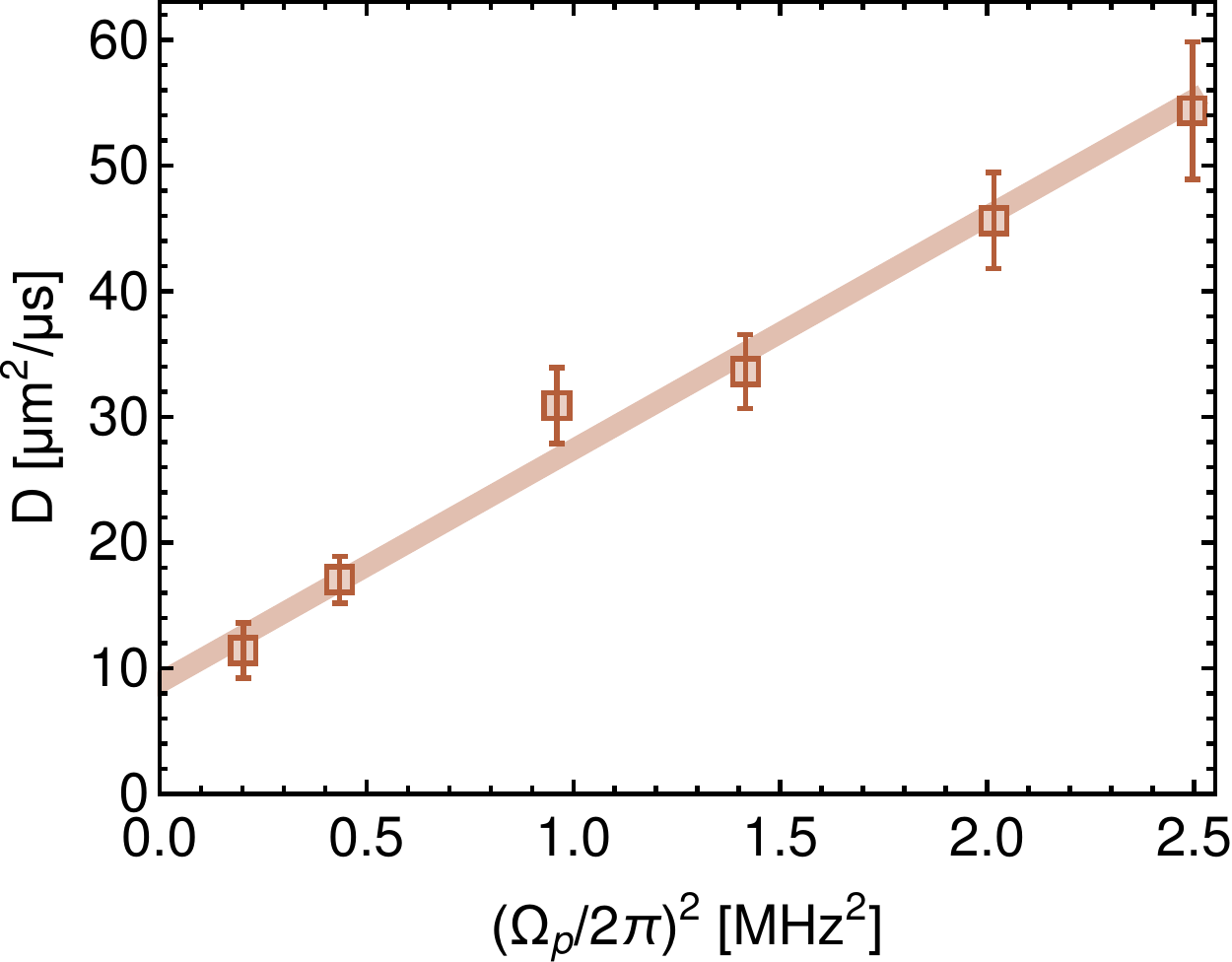}
 \caption{(Color online) Dependence of the diffusion coefficient on the dissipative medium set by the probe Rabi frequency $\Omega_p$. The data follow a linear dependence of the diffusion rate with probe intensity.}
 \label{fig3}
\end{figure}

Crucially, the observed diffusion rate can be controlled by the EIT laser fields used to create the dissipative medium. As the intensity of the probe laser field is varied we anticipate two effects: firstly the density of hopping partners is modified due to the dependence of the Rydberg $ns$ state admixture on the ratio of probe- and coupling laser intensities. Second, the rate of decoherence induced by the continuous measurement process changes with the photon scattering rate. Thus, under continuous observation and in the weak probe limit the classical hopping rate and the diffusion rate should scale proportional to the the probe laser intensity~\cite{guenter2013,schempp2015}.
To test this expectation, we measured the diffusion rate for a fixed exposure time of $T_\mathrm{exp}=21\,\mathrm{\mu s}$ as a function of probe intensity, expressed in terms of the square of the probe Rabi frequency in Fig.~\ref{fig3}. The data are consistent with the expected proportional intensity dependence. However, in contrast to the assumption of simple diffusive transport, we find a minimum diffusion rate of $9(2)~\mathrm{\mu m^2/\mu s}$ (obtained for $\Omega_p\rightarrow 0$). This rate is too large to be explained by thermal motion of the impurity atoms and is likely an indication of a different transport mechanism when the photon scattering rate becomes small compared to the coherent hopping rate.

\noindent\emph{Experiment (2) - Pulsed observation}. We now explore the transport properties of the Rydberg medium in the absence of the dissipative environment. As in the previous experiments we prepare impurities in state $|n's\rangle$. This is then followed by a $2~\mu$s probe laser pulse which triggers the fission process and establishes the background medium with the same density of $|ns\rangle$ Rydberg excitations as in experiment (1). We confirm that the initial conditions for experiment (1) and (2) are equivalent by the fact that the width of the impurity distribution after this pulse measured with a $5\,\mathrm{\mu s}$ exposure time matches the continuous observation case for $T_\mathrm{exp}=7\,\mathrm{\mu s}$ (Fig.~\ref{fig2}c) and that the number of $ns$ atoms in the gas meansured by field ionization detection remains unchanged during and after the excitation pulse. After this preparation phase we allow the system to evolve in the absence of probe light for a variable time $T_\mathrm{evol}$. This is then followed by a final 5\,$\mu$s long probe pulse which we use to record the spatial distribution of the impurity atoms. 

Figure\,\ref{fig2}c compares the case of pulsed observation (circles) to that of the continuous observation (squares) as a function of the integrated evolution and probe laser exposure times. For pulsed observation we observe a time dependence which is consistent with a total freezing of the hopping dynamics beyond what can be explained by classical diffusion during the preparation and imaging pulses (Fig.~\ref{fig2}b,c). This is highly suggestive of localization-like behavior on the dipolar hopping dynamics due to the random positions of the background gas atoms. Strictly speaking however, true localization (in the Anderson sense) should not be expected given the long-range nature of the dipolar interactions.

\begin{figure}[t!]
\includegraphics[width=0.95\columnwidth]{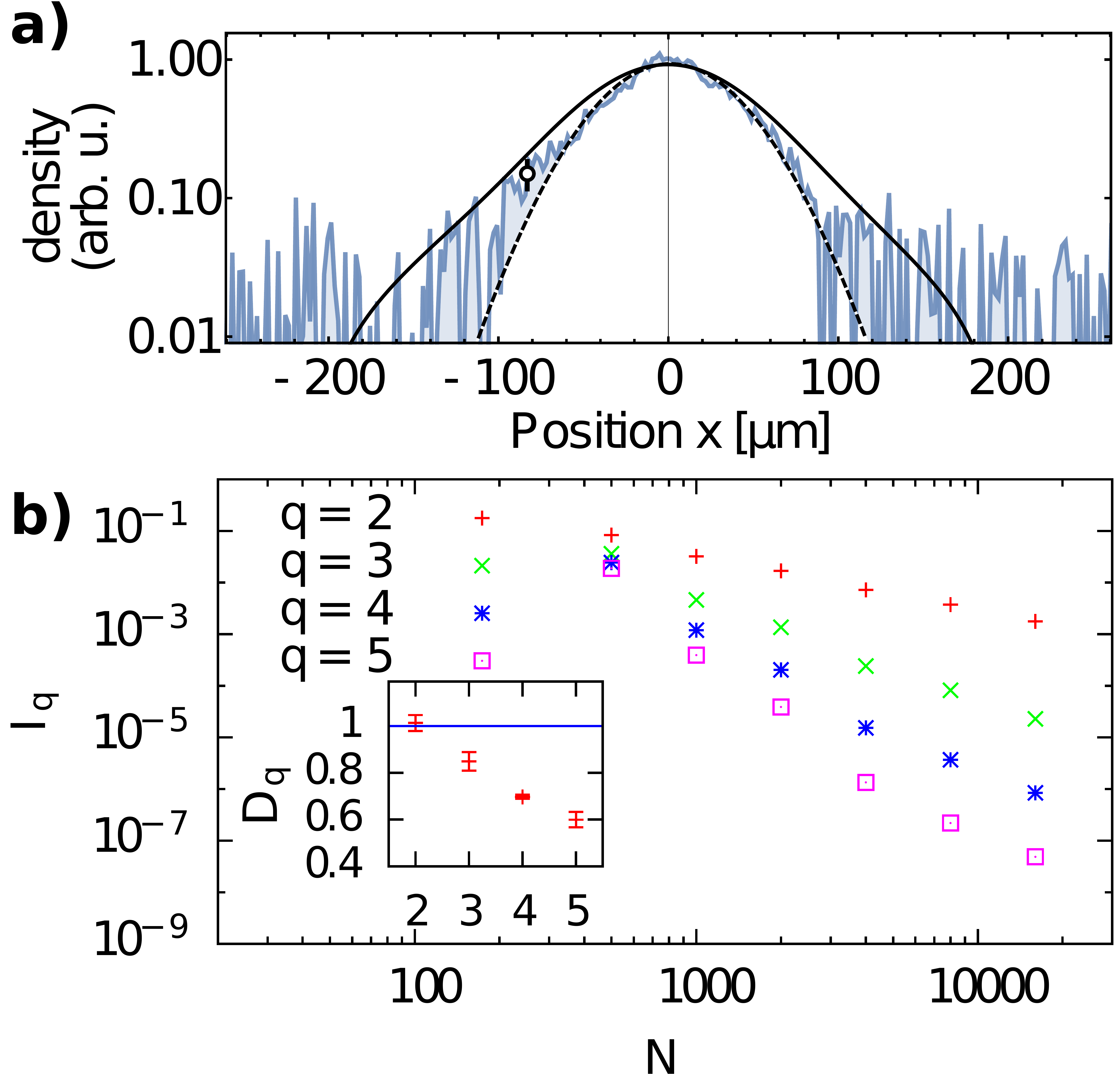}
\caption{Comparison to numerical simulations assuming purely coherent evolution under dipolar exchange interactions. (a) Average density profile obtained from the pulsed observation datasets. The dashed black line shows a Gaussian distribution assuming a frozen system with classical diffusion during the combined preparation and probe periods using the diffusion coefficient extracted from Fig.~\ref{fig2}c. The solid black line is the result of numerical simulations described in the text. A single datapoint is shown on the left-hand side in black with representative error bars corresponding to one standard error of the measurements. (b) Finite size scaling of the moments $I_q$ of the wave function. The inset shows the fractal dimension $D_q$ obtained from a power law fit for large system sizes. The solid line corresponds to $D_q = 1$ for ergodic states.}
\label{fig4}
\end{figure}

To test if the data is in fact consistent with the system entering a localized state we perform a careful analysis of the stationary spatial distribution of impurity atoms. Localization is often accompanied by the presence of exponentially decaying tails, while classical diffusion predicts asymptotically Gaussian distributions. Figure~\ref{fig4}a shows the measured distribution on a log-linear scale. To improve the signal-to-noise ratio in the wings we subtract the background absorption using the methods described in Ref.~\cite{Ockeloen2010} and average over all the data for $T>7\,\mathrm{\mu s}$ where the variance appears constant, corresponding to $1600$ runs of the experiment. The center of the distribution can be described by a Gaussian function according to the previously measured diffusion law given the total probe exposure time (dashed black line in the plot). However we observe significant deviations from a Gaussian in the wings, where the data are better described by exponential tails which would appear as straight lines on a logarithmic scale. The emergence of such exponentially decaying density profiles is a hallmark of Anderson-like localization phenomena~\cite{Lee1985}.

To explain these observations, we perform numerical simulations assuming purely coherent evolution under the dipolar hopping Hamiltonian. For convenience we limit ourselves to the single impurity subspace, i.e., equivalent to neglecting interactions between the impurities. In our simulations we randomly place up to 16\,000 background Rydberg atoms in a three-dimensional elongated volume that is comparable to the Gaussian cloud shape from the experiment, including a constraint to satisfy Rydberg blockade effects between nearby atoms. The simulations start with the impurity excitation localized on a central atom which is then evolved according to the time dependent Schr\"odinger equation. Finally we extract the spatial probability distribution for the impurity excitation as a function of time, averaging over many disorder realizations. Generally, the simulations show a period of rapid spreading over a time scale of approximately $1\,\mathrm{\mu s}$ which then freezes leaving leaving an approximately exponential probability distribution extending to $\pm 100~\mathrm{\mu m}$. To compare to the experimental data we account for diffusive transport during the preparation and imaging stages of the experiment by convolving the spatial distribution obtained from the coherent dynamics with a Gaussian function corresponding to the solution of the respective diffusion equation. The results are in good agreement with the experimental observations (c.f. solid black line in Fig.~\ref{fig4}a), in particular reproducing the apparent absence of spreading and the exponential wings, evidencing the localization properties of the system. The numerical simulations are in better agreement with the data than the diffusion model alone (dashed line in Fig.~\ref{fig4}a). The remaining discrepancies between the data and the theoretical calculations, such as the slightly larger localization length reflected by the broader exponential wings seen in theory, might hint at effects in the experiment beyond the coherent dipolar hopping dynamics of independent impurities.

Within the numerical simulations, we can now investigate in the cause for the absence of spreading observed in the experiment. We perform a finite size scaling of the higher order moments of the single particle eigenstates,
\begin{equation}
I_q = \sum\limits_j |\psi(j)|^{2q},
\end{equation}
where the index $j$ runs over all the sites of the system. Here, we focus on eigenstates at the center of the band, as these play the dominant role in the experimental setting. For the finite size scaling, we keep the density and aspect ratio at fixed values. Generically, $I_q$ will scale as a power law $I_q \sim N^{D_q(q-1)}$ with $N$ being the number of sites~\cite{DeLuca2014}. The fractal dimension $D_q$ can be used to classify the transport behavior for the system: $D_q = 0$ signals the presence of Anderson localization, while states with finite $D_q$ are extended. However, deviations from $D_q = 1$ signal the presence of non-ergodic extended states \cite{DeLuca2014}. 

Indeed, from the finite size scaling of $I_q$, we find a significant reduction of the fractal dimension $D_q$ for $q > 2$, see Fig.~\ref{fig4}b. This
breaking of ergodicity explains why the system appears to remain stuck in the center of the cloud despite the long-range hopping. We consider this as evidence that the presence of non-ergodic extended states causes the dramatic reduction of the spreading as compared to the diffusive case. We note that this finding is in agreement with other simulations for dipolar transport in disordered lattice models~\cite{Deng2016}.

In conclusion, we have studied the dipolar energy transport dynamics in Rydberg gases in two different scenarios: 1) continuous observation resulting in transport through a dissipative medium and 2) pulsed observation during which the impurities are free to evolve by purely coherent dynamics. In the first case we observe dynamics characteristic of classical diffusion, with a diffusion coefficient that can be controlled by more than a factor of five by varying the intensity of the probe laser field. In the case of pulsed observation we find that the spreading of impurities is effectively frozen which we attribute to the system entering a non-ergodic phase. These experiments establish Rydberg excitations as a unique platform to study quantum transport including novel regimes arising as a consequence of long-range interactions. Our experiments, combined with the high degree of flexibility afforded by ultracold gases to control the spatial geometry, types of disorder, range of interactions and degree of coherence, paves the way for future studies addressing important outstanding questions on many-body localization, non-equilibrium phase transitions, and the dynamics of non-ergodic extended states.

We thank J. Evers, X. Deng, G. Pupillo and L. Santos for fruitful discussion as well as G. G\"unter and V. Gavryusev for contributions to the experimental apparatus. This work is part of and supported by the DFG Collaborative Research Centres ''SFB 1225 (ISOQUANT)'' and ''SFB 1227 (DQ-mat)'', the Priority Programme DFG SPP 1929 GiRyd, the Heidelberg Center for Quantum Dynamics, the European Union H2020 FET Proactive project RySQ (grant N. 640378) and the `Investissements d'Avenir' programme through the Excellence Initiative of the University of Strasbourg (IdEx). S. Whitlock was partially supported by the University of Strasbourg Institute for Advanced Study (USIAS) and H. Weimer acknowledges support by the Volkswagen Foundation.

\bibliography{transport}

\end{document}